# Interfacial Ferroelectricity by van der Waals Sliding

M. Vizner Stern,[1] Y. Waschitz,[1] W. Cao,[2] I. Nevo,[1] K. Watanabe,[3] T. Taniguchi,[4]
E. Sela,[1] M. Urbakh,[2] O. Hod,[2] M. Ben Shalom[1]*

[1] Raymond and Beverly Sackler School of Physics and Astronomy, Tel-Aviv University, Israel

[2] Department of Physical Chemistry, School of Chemistry, The Raymond and Beverly Sackler Faculty of Exact Sciences and The Sackler Center for Computational Molecular and Materials Science, Tel Aviv University, Tel Aviv 6997801, Israel

[3] Research Center for Functional Materials, National Institute for Materials Science, 1-1 Namiki, Tsukuba 305-0044, Japan

[4] International Center for Materials Nanoarchitectonics, National Institute for Materials Science, 1-1 Namiki, Tsukuba 305-0044, Japan

**Despite their ionic nature, many layered diatomic crystals avoid internal electric polarization by forming a centrosymmetric lattice at their optimal anti-parallel van-der-Waals stacking. Here, we report a stable ferroelectric order emerging at the interface between two naturally-grown flakes of hexagonal-boron-nitride, which are stacked together in a metastable non-centrosymmetric parallel orientation. We observe alternating domains of inverted normal polarization, caused by a lateral shift of one lattice site between the domains. Reversible polarization switching coupled to lateral sliding is achieved by scanning a biased tip above the surface. Our calculations trace the origin of the phenomenon to a subtle interplay between charge redistribution and ionic displacement, and our minimal cohesion model predicts further venues to explore the unique "slidetronics" switching.**

The ability to locally switch a confined electrical polarization is a key requirement in modern technologies, where storing and retrieving a large volume of information is vital [1]. Today, after decades of intense research, it is possible to squeeze a tera ($10^{12}$) of polarized islands in a cm$^2$ chip [2] composed of high-quality three-dimensional (3D) crystals [3]. The need to further reduce the dimensions of individually polarized domains, however, from the ~ 100 nm$^2$ scale towards the atomic scale is rising [4]. The main challenges involve long-range dipole interactions which tend to couple the individual domain polarization orientations [5]. Likewise, surface effects and external strains that are difficult to control become dominant once the surface-to-volume ratio increases [6]. Venues to overcome the abovementioned difficulties become straightforward when considering 2D crystals. In particular, in layered materials such as hexagonal-boron-nitride (*h*-BN) and transition-metal-dichalcogenides (TMD), the bulk volume can be reduced to the ultimate atomic-thickness limit whereas the crystalline surface remains intact [7]. On the other hand, it is rare to find a spontaneous net electric polarization in 2D that is sufficiently large to read and write under ambient conditions [8–10]. Furthermore, in common *h*-BN and TMD crystals, polarization is eliminated by a centrosymmetric van der Waals (vdW) structure which is lower in energy than other stacking configurations. In the present study, we break this symmetry by controlling the twist-angle between two *h*-BN flakes, and find an array of permanent and switchable polarization domains at their interface. Oriented normal to the plane, the polarization amplitude is in good agreement with a first-principle prediction for a two-layers system only [11], and with our finite system calculations.

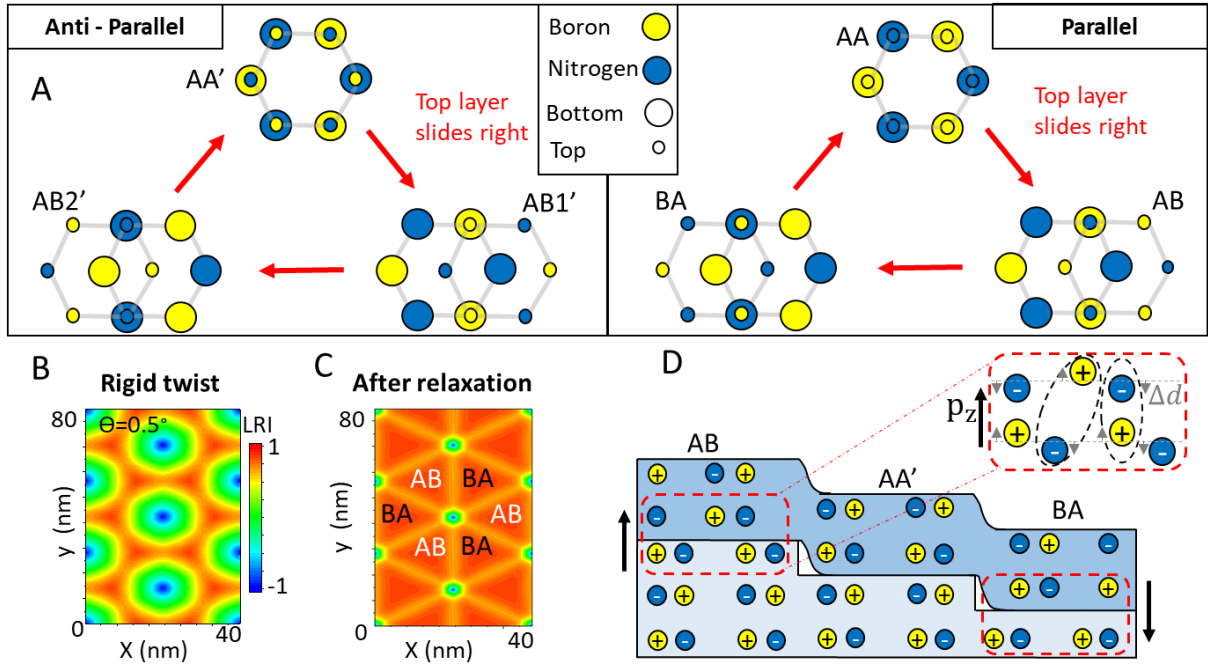

**Fig. 1. High symmetry interlayer stacking configurations.** (A) Top view illustration of two layers. For clarity, atoms of the top layer are represented by small circles. For either group of parallel / anti-parallel twist orientations, a relative lateral shift by one lattice spacing results in a cyclic switching between three high-symmetry stacking configurations. (B) Local-registry index (LRI, defined in SM) map of two nearly parallel rigid layers with a twist angle of 0.5°. Blue regions correspond to AA stacking, whereas AB/BA stacking appear in orange (LRI = 0.86). (C) Calculated LRI map after geometry relaxation of the structure presented in panel (B). Large domains of uniform untwisted AB/BA stacking appear, on the expense of the pre-optimized AA regions. The twist is accumulated in smaller AA–like regions and in the ~ 10 nm wide incommensurate domain walls (bright lines), see Fig. S1 and SM for further discussion. (D) Cross-sectional illustration of two few-layered flakes (blue and light blue regions marked the top and bottom flakes, respectively) of naturally grown *h*-BN (AA'), which are stacked with no twist. Plus (minus) signs mark boron (nitrogen) sites. A topographical step of a single layer switches between parallel and anti-parallel stacking orientations at the interface between the two flakes. Vertical charge displacements and the resulting net polarization $P_z$ are marked by arrows.

To identify which stacking modes can carry polarization, we present in Figure 1A six different high symmetry configurations of bilayer *h*-BN. The stacking configurations are divided into two groups termed as "parallel" and "anti-parallel" twist orientations [12], where within each group, a relative lateral shift by one interatomic distance switches the stacking configuration in a cyclic manner. Typically, the crystal grows in the optimal anti-parallel (AA') configuration with full overlap between nitrogen (boron) atoms of one layer and boron (nitrogen) atoms of the adjacent layer [13]. In the parallel twist-orientation, however, the fully eclipsed configuration (AA) is unstable since it forces pairs of bulky nitrogen atoms atop each other resulting in increased steric repulsion [14]. Instead, a lateral interlayer shift occurs to a metastable AB stacking with only half of the atoms overlapping, whereas the other half are facing empty centers of the hexagons in the opposite layer [15,16]. Note that the AB and BA stacking form equivalent lattice structures (only flipped), and that all depicted anti-parallel configurations (AA', AB1', AB2') are symmetric under spatial inversion.

To explore these different configurations, we artificially stamped two exfoliated *h*-BN flakes on top of each other, each consisting of a few AA' stacked layers, with a minute twist-angle between the otherwise parallel interfacial layers, see details in supplementary materials (SM). The small twist imposes interlayer translations that evolve continuously and form a Moiré pattern owing to the underlying crystal periodicity (Fig. 1B). In this rigid lattice picture, the three nearly commensurate stacking configurations (AB, BA, AA) appear at adjacent positions in space. Notably, this picture breaks for a sufficiently small twist angle as a result of structural relaxation processes as shown by our molecular dynamics calculation based on dedicated interlayer potentials (Fig. 1C, SM) [17]. Instead, the system divides into large domains of reconstructed commensurate AB and BA stackings separated by sharp incommensurate domain walls that accommodate the global twist (see Fig. S1A,B) [18–21]. Notably, near the center of the extended commensurate domains, perfectly aligned configurations are obtained with no interlayer twist. In the experiments, we also introduce a topographic step at the interface between the flakes. A step thickness of an odd number of layers guarantees anti-parallel stacking (AA', AB1' or AB2') on one side, and parallel stacking (AA, AB or BA) on its other side (Fig. 1D). Thus, one can compare all possible configurations at adjacent locations in space.

To measure variations in the electrical potential, $V_{KP}$, at surface regions of different stacking configurations, we place the *h*-BN sandwich on a conducting substrate (graphite or gold), and scan an atomic force microscope (AFM) operated in a Kelvin probe mode (KPFM), see Fig 2A and SM. The potential map above the various stacking configurations is shown in Fig 2B. We find clear domains (black and white) of constant $V_{KP}$, extending over areas of several $\mu m^2$, which are separated by narrow domain-walls. Gray areas of average potential are observed above: (i) positions where only one *h*-BN flake exists (outside the dashed yellow line); (ii) above two flakes but beyond the topographic step marked by dashed red line in Fig. 2B (and topography map Fig. S2) as expected; and (iii) beyond topographic folds that can further modify the interlayer twist angle (dashed green line). These findings confirm that white and black domains correspond to AB and BA interfacial stacking that host a permanent out-of-plane electric polarization. Such polarization is not observed at the other side of the step, where centrosymmetric AA', AB1' AB2' configurations are obtained, or at the AA configuration expected at domain-wall crossings (see blue dots in Fig. 1C). Sufficiently far from the domain-walls a constant potential is observed, allowing us to measure the potential difference between the AB and BA domains giving $\Delta V_{KP} = 100 \pm 10$ mV, as shown in Fig. 2C. We note that similar values for $\Delta V_{KP}$ are obtained for several samples (Fig. S2), regardless of the substrate identity (Si, graphite, or gold), the type and height of the AFM tip during the scan, and the thickness of the top *h*-BN flake (for flakes thicker than 1 nm). These findings confirm that $\Delta V_{KP}$ is an independent measure of the intrinsic polarization of the system that, in turn, is confined within a few interfacial layers.

While our force field calculations for slightly twisted bilayer *h*-BN show a uniform triangular lattice of alternating AB and BA stacked domains (Fig. 1C), in the experiment we observe large variations in their lateral dimensions and shape. This indicates minute deviations in the local twist, which are unavoidable in the case of small twist angles [18–21]. Specifically, the ~1 $\mu m^2$ domains in the left part of Fig. 2B correspond to a global twist of less than 0.01 degree [22]. We also note that any external perturbation to the structure, caused either by transferring it to a polymer, heating, or directly pressing it with the AFM tip, usually resulted in a further increase in domain size. In a few cases, high-temperature annealing resulted in a global reorientation to a single domain flake, many micrometers in dimensions. This behavior confirms the metastable nature of the AB / BA stacking mode, as well as the possible superlubric nature of the interface [23,24]. On the other extreme, much smaller domains are observed in the top right-hand section of Fig. 2B. The smallest triangle edge we could identify over many similar flakes was 60 nm in length, which corresponds to a twist-angle of 0.24°. We therefore conclude that below this angle atomic reconstruction to create untwisted domains is energetically favored. Naturally, this constitutes a lower bound on the maximal angle for domain formation as smaller domains below our experimental resolution may form at larger twist angles.

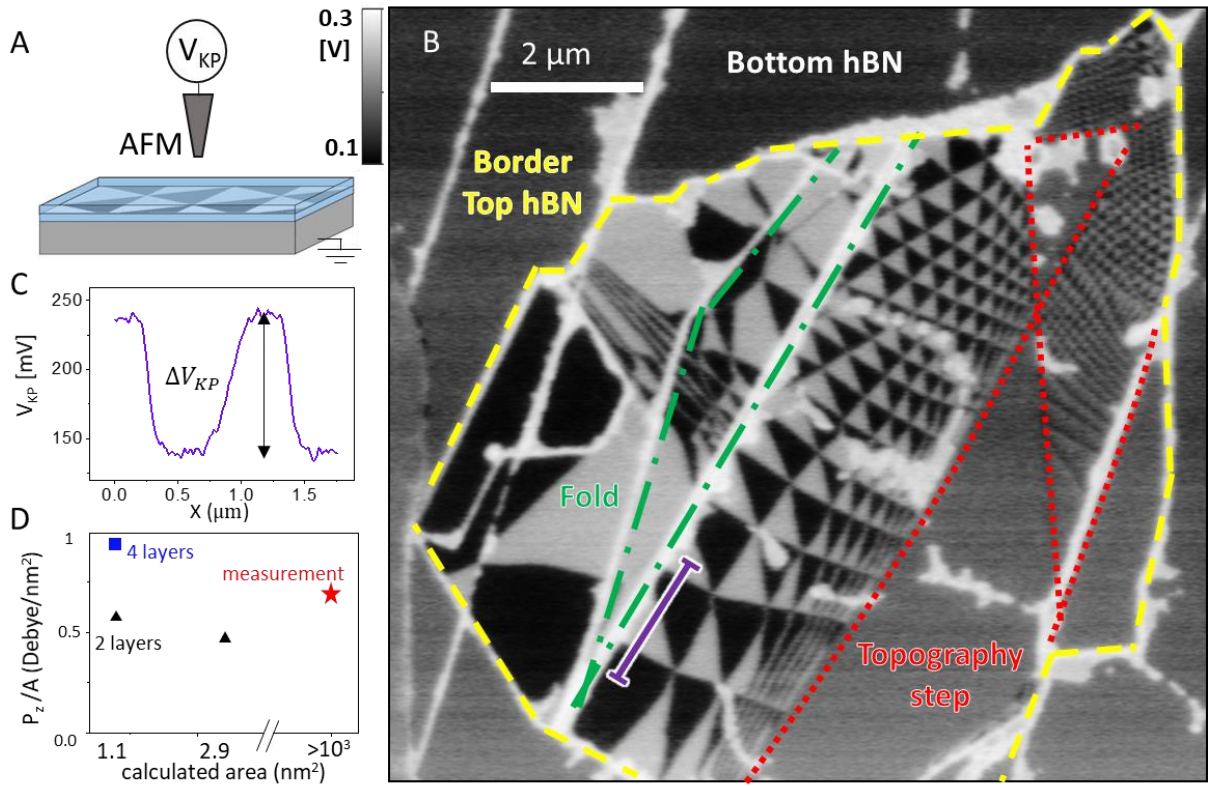

**Fig. 2. Direct measurement of interfacial polarization.** (A) Illustration of the experimental setup. An atomic force microscope is operated in Kelvin-probe mode to measure the local potential modulation, $V_{KP}$, at the surface of two 3 nm thick *h*-BN flakes, which are stacked with a minute twist angle. (B) $V_{KP}$ map showing oppositely-polarized domains of AB/BA stacking (black and white), ranging in area between ~0.01 and 1 μm² and separated by sharp domain-walls. (C) Surface potential along the purple line marked in (B). (D) DFT calculations of the polarization $P_z$ per unit area in two-layer (black triangles) and four-layer (blue square) finite *h*-BN stacks of different lateral dimensions. The out-of-plane polarization estimated based on the measured potential is shown on the right (red star).

To trace the microscopic origin of the measured polarization we performed a set of density functional theory (DFT) calculations on finite bilayer and quad-layer *h*-BN flakes (see SM). For the finite bilayer calculations, we constructed two model systems, where hydrogen passivated *h*-BN flakes of either 1 nm² or 3 nm² surface area are stacked in the AB stacking mode (see Fig. S3). The calculated polarizations per unit area were 0.55, 0.45 Debye/nm², respectively (see black triangles in Fig. 2D), pointing perpendicular to the interface. The fact that the calculated polarization values for the two finite flakes are similar indicates that edge effects have a relatively small contribution. This is further supported by the smaller effective charges residing on the edge atoms in both bilayer systems (see Fig. S3). To model a finite four-layer system we added two

AA' stacked layers at the two sides of the AB stacked interface. To avoid excessive computational burden, we limited this calculation to a surface area of 1 nm². The resulting polarization obtained was 0.91 Debye/nm² (see Fig. 2D, blue square), larger than that obtained for the bilayers. We attribute this to remnant edge contributions in the finite-size structure (see SM). Comparison with the experimental observations can be performed by adopting a parallel plate capacitor model that translates the measured potential drop, $\Delta V_{KP}$, to the effective system polarization via $P_z = \frac{\Delta V_{KP}}{2}\varepsilon_0\varepsilon_r A$, where $A$ is the contact surface area and $\varepsilon_0$ and $\varepsilon_r = 3$ are the vacuum permittivity and the relative normal permittivity of single-layered $h$-BN [25], respectively. The resulting polarization per unit area is $0.66\ Debye/nm^2$ (red star in Fig. 2D) in reasonable quantitative agreement with the calculated values.

It is instructive to further translate the measured potential difference into intra-layer displacements in a simplistic point-like charges model (see $\Delta d$ in Fig. 1D), where each atom is allowed to displace from its layer's basal plane in the vertical direction. Note that using the lattice site density of $n = 37\ nm^{-2}$, and the on-site charge value, $q \sim e/2$, for single-layered $h$-BN (26), our measured $\Delta V_{KP}$ gives out-of-plane atomic displacement of the order of $\Delta d = \Delta V_{KP}\varepsilon_0\varepsilon_r/4nq \sim 2 \times 10^{-3}$Å, which is significantly smaller than the inter-lattice (1.44 Å) and interlayer (3.30 Å) spacing. This implies that the polarization is determined by a delicate competition between the various interlayer interaction components and charge redistribution. Intuitively, we expect the vdW attraction to vertically compress the non-overlapping interfacial sites (diagonal dashed ellipse in Fig.1D) closer together than the overlapping sites (vertical dashed ellipse in Fig.1D) which are more prone to Pauli repulsion. This direction of motion, for example, reduces the average interlayer separation and favors Bernal (AB like) stacking in graphite over the AA configuration [28]. In $h$-BN, however, the ionic nature of the two lattice sites matters [12,29,30], making the fully eclipsed AA' stacking more stable [13]. Hence, imposing a polar AB interface, as in our case, may favor overlapping sites of opposite charges to come closer together than the non-overlapping pairs and the polarization to point in the opposite direction.

To quantify these arguments, we present a reduced classical bilayer model that captures the intricate balance between Pauli, vdW, and Coulomb interatomic interactions at different stacking modes. In our model (see SM), the interfacial energy $E = \frac{1}{2}\sum_{i,j}\left[4\varepsilon\left(\left(\frac{\sigma}{r_{ij}}\right)^{12} - \left(\frac{\sigma}{r_{ij}}\right)^6\right) + \frac{q_i q_j}{r_{ij}}\right]$, includes a Lennard-Jones (LJ) potential characterized by the cohesive energy, $\varepsilon$, and the interlayer spacing scale $\sigma$, and Coulomb interactions between the dimensionless partial atomic charges on the boron and nitrogen sites $q = \pm q_i/e$. Although neglecting any charge transfer processes between the layers that are explicitly taken into account in our DFT calculations, this model captures both the magnitude and orientation of the polarization by adjusting the ratio between $\varepsilon$ and Coulomb scales $\propto q^2/\sigma$ (see Fig S4). We note that our detailed DFT calculations indicate that in bilayer $h$-BN the net polarization is oriented as marked by the arrows in Fig. 1D.

The emergence of permanent polarization observed in separated domains, whose dimensions can be tuned by the twist-angle, each exhibiting a distinct and stable potential, opens the door for novel applications. To that end, however, one should identify additional ways to control the local orientation beyond the twisting mechanism. Note that polarization inversion can be achieved by reversible switching between AB and BA configurations which, in turn, can be realized by relative lateral translation by one atomic spacing (1.44 Å) as illustrated in Fig. 1A. Similar sliding between different stacking configurations was recently demonstrated in multilayer graphene. It was shown that both mechanical [31] or electric perturbations [32] can push domain-walls, practically modifying the local stacking. In the present *h*-BN interface, however, the polar switching calls for a preferred up or down orientation which can be predetermined by the user. To obtain such a spatially resolved control we scanned a biased tip above an individual domain to induce a local electric field normal to the interface. The polarization images before and after the biased scans are presented in Fig 3. We observe a redistribution of domain-walls to orient the local polarization with the electric field under the biased tip. For example, after scanning a negatively biased tip above the region marked by the blue square, we observed a large white domain due to the motion of the walls marked by dashed red (green) line before (after) the scan. A successive scan by a positively biased tip resulted in practically complete domain polarization flipping. Hence, by selecting negative or positive bias to the tip it is possible to determine the polarization orientation of the underlying domain. We note that domain-wall motion is observed for electric field values exciding ~ 0.3 V/nm, and when operating the biased scan in a pin-point mode (see SM). Similar switching behavior was attained above different domains within the same interface and for several measured structures (see Fig S5).

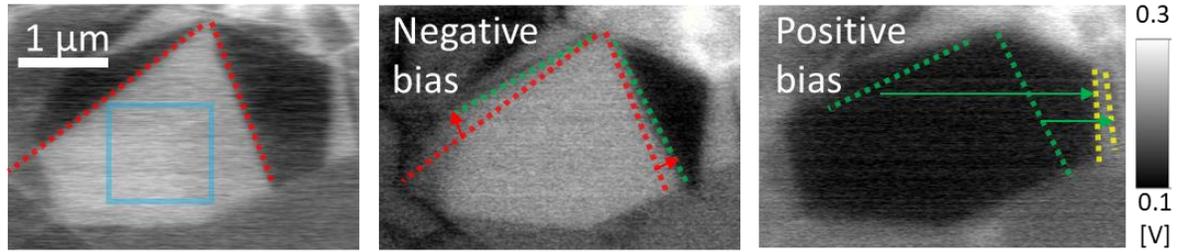

**Fig. 3. Dynamic flipping of polarization orientation by domain-wall sliding.** Kelvin-probe maps measured consequently from left to right above a particular flake location showing domains of up (white) and down (black) polarizations. The middle image was taken after biasing the tip by a fixed DC voltage of -20 Volts and scanning it above the blue square region shown on the left-hand image. Then the tip was biased by 10 Volt and scanned again over the same region before taking the right-hand image. Consequent domain-walls positions are marked by dashed red, green, and yellow lines. Larger white (black) domains appear after positive (negative) bias scans as a result of domain-wall motions. Note that the number of domain-walls is apparently not altered.

Our results therefore demonstrate that the broken symmetry at the interface of parallel-stacked *h*-BN flakes gives rise to an out-of-plane two-dimensional polarization confined within a few interfacial layers, that can be locally detected and controlled. While the *h*-BN system, with its only two different light atoms, offers a comfortable experimental and computational testbed and allows for intuitive interpretations, first-principle analysis predicts similar phenomena to occur in other more complex bi-atomic vdW crystals [11], such as various TMD [33,34]. Notably, the origin of the polarization and the inversion mechanism discovered herein are fundamentally different from the common deformations of tightly-bonded atoms in non-centrosymmetric 3D bulk crystals. The "slidetronics" switching involves lateral domain-wall motion in a weakly-coupled interface under ambient conditions. The sensitivity of the system to the delicate interplay between van der Waals attraction, Pauli repulsion, Coulomb interactions, and charge redistribution implies that external stimuli such as pressure, temperature, and/or electric fields may be used to control the polarization, thus offering many opportunities for future research.

**Acknowledgments:**
We thank Yoav Lahini for useful discussions and Neta Ravid for laboratory support. W.C. acknowledges the financial support of the IASH and the Sackler Center for Computational Molecular and Materials Science at Tel Aviv Univeristy. Growth of hBN was supported by the Elemental Strategy Initiative conducted by the MEXT, Japan, Grant Number JPMXP 0112101001, JSPS KAKENHI Grant Number JP20H00354 and the CREST (JPMJCR15F3), JST. E.S. acknowledges support from ARO (W911NF-20-1-0013), the Israel Science Foundation grant number 154/19 and US-Israel Binational Science Foundation (Grant No. 2016255). M.U. acknowledges financial support of the Israel Science Foundation, Grant 1141/18 and the ISF-NSFC joint Grant 3191/19. O.H. is grateful for the generous financial support of the Israel Science Foundation under Grant 1586/17, the Naomi Foundation for generous financial support via the 2017 Kadar Award, and the Ministry of Science and Technology of Israel under project number 3-16244. M.B.S. acknowledge funding by the European Research Council (ERC) under the European Union's Horizon 2020 research and innovation programme (grant agreement No. 852925), the Israel Science Foundation Grant No. 1652/18, and the Israel Ministry of Science & Technology project number 3-15619 (Meta-Materials). O.H. and M.B.S acknowledge the Center for Nanoscience and Nanotechnology of Tel-Aviv University.

# Materials and methods

### a) Device fabrication

*h*-BN flakes of various thicknesses (1-5 nm) were exfoliated onto a SiO$_2$ surface. A particular flake is selected to have several topographic steps of a few-layers thickness. The flake is ripped off into two pieces which are stacked together by a polymer stamp [35]. During the stamping processes we make sure to minimize any twist orientation. Finally, the two-flakes-sandwich is placed on a conducting graphite flake or alternatively on a gold substrate, using the same dry transfer method.

### b) AFM scans

Topography and Kelvin probe force microscopy (KPFM) measurements are acquired simultaneously (Fig S2), using Park System NX10 AFM in a non-contact scanning mode. The electrostatic signal is measured in the first harmonic using a built-in lock in amplifier. We use Multi-75g and PPP-EFM n-doped tips with conductive coating. The mechanical resonance frequency of the tips is 75 kHz, and the force constant is 3 N/m. The cantilever is oscillated mechanically with an amplitude of ~ 20 nm. The cantilever is also excited with an AC voltage to perform KPFM measurements as described below, with amplitude of 3-6 V and frequency of 17 kHz. The DC voltage is controlled by a servo motor to obtain the surface potential measurements. The images are acquired using Park SmartScan software and the data analysis is performed with Gwyddion program.

To switch the domain orientation biased scans are performed in a pin-point mode. Here the tip approaches the surface vertically at each pixel in the scanned area. The estimated maximum force during this approach is 50nN. This mode minimizes lateral forces between the tip and the surface.

### c) Kelvin-probe surface potential measurements

The AFM tip and the sample are treated as a parallel-plate capacitor model [36]. The charge induced on the tip and the substrate is affected by the voltage applied between them, and potential drops related to the sample.

The applied voltage on the tip consists of DC and AC components. The total voltage is given by:

$$V = V_{DC} + V_{AC}\sin(\omega t) + V_{CPD}$$

where $V_{CPD}$ is the contact potential difference, which originates from the different work function of the tip and the substrate. The force acting on the tip is: $\quad F = \frac{A}{2\epsilon}\rho^2,$ where $A$ is the effective area of the capacitor, $\epsilon$ is the dielectric constant and $\rho$ is the two-dimensional charge density. The latter can be extracted from: $V = \frac{\rho}{\epsilon}d + V_{int}$, where $d$ is the distance between the plates, and $V_{int}$ is the voltage drop at the *h*-BN interface. This claim holds assuming the sample is neutral and the field outside the sample from the charges distribution in the sample is zero. After inserting it in the force equation, we get the first harmonic of the force:

$$F(\omega) \propto (V_{int} - V_{DC} - V_{CPD})V_{AC}\sin(\omega t)$$

It vanishes for $V_{DC} = V_{int} - V_{CPD}$. The main principle of KPFM is to apply a DC voltage that nullifies the first harmonic, so $V_{int}$ signal can be extracted from variation in the KPFM signal, $V_{DC}$, at different lateral positions above the surface.

### d) Model system and classical force-field calculations

To study the structural properties of twisted $h$-BN interfaces we constructed a model system consisting of two $h$-BN layers with an interlayer misfit angle of ~0.5°. To mimic the experimental scenario, a laterally periodic supercell was constructed with a triangular lattice of periodicity $L = |n\vec{a}_1 + m\vec{a}_2|$, where the primitive lattice vectors are given by $\vec{a}_1 = a_{hBN}(\sqrt{3}, 0)$ and $\vec{a}_2 = \frac{a_{hBN}}{2}(\sqrt{3}, 3)$ and $a_{hBN} = 2.505$ Å based on the Tersoff potential equilibrium bond-length of $b_{BN} = 1.446$ Å. The indices $n = 195$ and $m = 1$ were chosen to fulfil the condition:

$$\cos(\theta) = \frac{2n^2 - m^2 + 2nm}{2(n^2 + m^2 + nm)}. \tag{S1}$$

The corresponding moiré pattern dimension is $L = \frac{b_{BN}}{\sqrt{2 - 2\cos(\theta)}} = 16.3$ nm. The parallelepiped supercell was then multiplicated to construct a rectangular supercell consisting of more than 300,000 atoms.

The structural properties of the twisted $h$-BN interface were calculated using the Tersoff [37] intra-layer potential in conjunction with the recently developed dedicated interlayer potential (ILP) [17]. We first optimized the geometry of the top layer atoms with fixed supercell size using the Fire algorithm and keeping the bottom layer rigid. This was followed by optimization of the supercell dimensions by the conjugate gradient (CG) algorithm while scaling the rigid bottom layer according to the simulation box size. This two-step energy minimization procedure was repeated for ten times, which is sufficient to obtain well converged results (see Fig. S1C). In each repetition, both minimization stages were terminated when the forces acting on each degree of freedom reduced below $10^{-3} eV/$Å. The system was further relaxed by Fire algorithm at a force tolerance of $10^{-4} eV/$Å. The optimized structure exhibits atomic reconstruction with distinct AB and BA stacking domains separated by domain walls (see Fig S1A).

### e) Local registry index analysis

The local registry index (LRI) (Fig. 1D) is a method introduced to quantify the degree of local interfacial registry matching at rigid material interfaces [38,39]. The idea is to assign a number between -1 and 1 to each atom in the layer, signifying whether it resides in an optimal or worse stacking region, respectively. To this end, a circle is associated with each atomic position in the two layers and the overlaps between circles of one layer and those of the adjacent layer are evaluated. For the case of $h$-BN three types of overlaps are considered, namely $S_i^{NN}$, $S_i^{NB} = S_i^{BN}$, and $S_i^{BB}$. Here, $S_i^{JK}$ signifies the overlap of the circle associated with atom $i$ of type $J$ in one layer with all circles associated with $K$ type atoms in the adjacent layer. The radius of the circles associated with B and N atoms is taken as $r_B = 0.15 b_{hBN}$, and $r_N = 0.5 b_{hBN}$, which provides good qualitative agreement between registry index maps and the sliding potential energy surfaces obtained from density functional theory calculations [38]. The LRI of atom $i$ is then defined as the average registry index of itself and its three nearest neighbors ($j, k, l$) within the entire layer, as follows:

$$LRI_i =$$

$$\frac{1}{3}\sum_{n=j,k,l}\frac{\left[(S_i^{NN}+S_n^{NN})-(S_i^{NN,opt}+S_n^{NN,opt})\right]+\left[(S_i^{BB}+S_n^{BB})-(S_i^{BB,opt}+S_n^{BB,opt})\right]-\left[(S_i^{NB}+S_n^{NB})-(S_i^{NB,opt}+S_n^{NB,opt})\right]}{[(S_i^{\ NN,worst}+S_n^{NN,worst})-(S_i^{NN,opt}+S_n^{NN,opt})]+[(S_i^{NN,worst}+S_n^{NN,worst})-(S_i^{NN,opt}+S_n^{NN,opt})]+[(S_i^{NB,worst}+S_n^{NB,worst})-(S_i^{NB,opt}+S_n^{NB,opt})]},\quad(S2)$$

where $S_i^{JK,opt}$ and $S_i^{JK,worst}$ are $S_i^{JK}$ evaluated at the optimal and worst local stacking modes, respectively (AA' and AA in the case of $h$-BN, respectively, see Fig 1A of the main text). The calculated $LRI_i$ is then transformed by $-(2LRI_i-1)$ to make it range be between [-1, 1]. With this, the LRI at an AA' (AA) stacked region is 1 (-1) respectively, and that of an AB stacked region is 0.86.

Plotting the LRI following geometry relaxation as discussed above (see main text Fig. 1D) we found an ordered array of AB and BA stacked domains separated by sharp domain walls. To estimate the width of the domain-wall region, we plotted a cross section of the out-of-plane height profile (see Fig. S1D) along the path marked by the dashed black line in Fig. S2B. A clear inverse correlation between the height map (red line) and the registry index (black line) is obtained. At the AB and BA stacked regions the interlayer distance is relatively constant at ~3.23 Å and correspondingly a relative constant value of LRI ~0.86 is obtained. At the center of the domain wall, the interlayer distance increases by ~0.05 Å and the LRI reduces to ~0.67, whereas at the domain wall crossings the interlayer distance reaches ~3.57 Å and the LRI drops to ~-0.93. Using a Gaussian fitting to the height profile near the domain wall we can estimate the domain wall width to be ~10 nm.

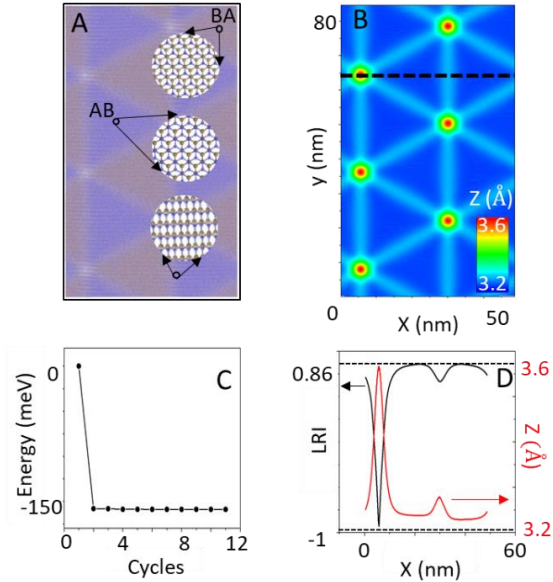

**Fig. S1**. **Geometric relaxation of Moiré pattern**; (A) Atomic configuration of a relaxed periodic twisted bilayer $h$-BN with a twisted angle of $\theta=0.5°$. The size of the entire model system is 48.9 nm × 84.7 nm. B and N atoms are colored by ochre and blue, respectively. Distinctly colored AB and BA domains are obtained since atoms of the upper layer hide those of the bottom layer that reside exactly below them. (B) Interlayer distance map for the relaxed twisted bilayer. (C) The energy variation during the minimization cycles applied to the model system appearing in panel (A) plotted relative to the initial energy. The first ten points represent cycles with force tolerance set to $10^{-3}$ eV/Å and the last point corresponds to the final minimization step with force tolerance of $10^{-4}$ eV/Å. (D) Interlayer distance (red) and local registry index (black) along the path marked by the black dashed line in panel B. The reference LRI values of the AA and AB/BA stacking modes are -1 and 0.86 (marked by the corresponding horizontal lines).

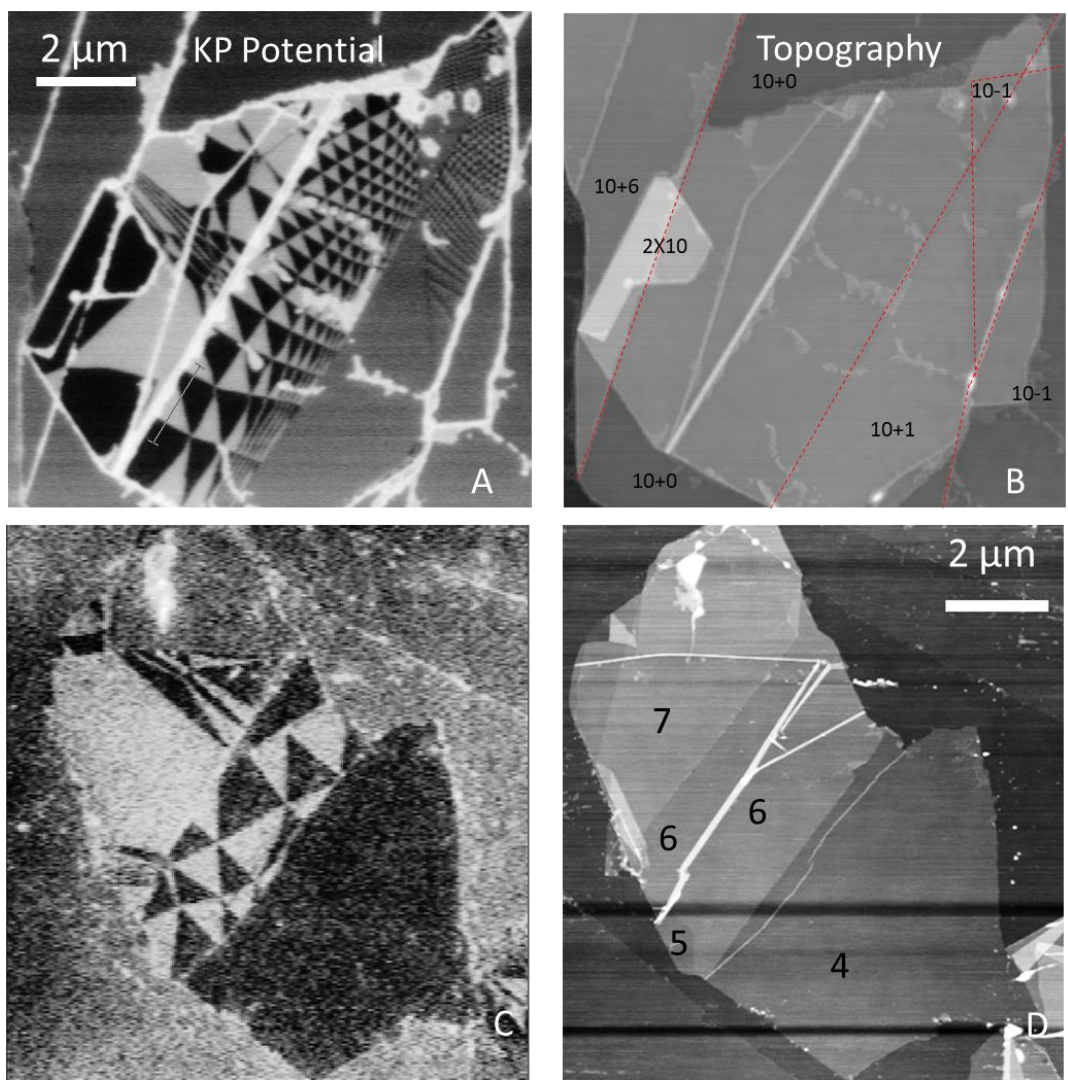

**Fig S2**. **Topography maps and additional samples**. (A, B) Surface potential and topography maps measured simultaneously on the structure presented in the main text (Fig 2B). The top *h*-BN flake thickness is uniform and includes 10 layers. Topography steps in the surface of the bottom flake are marked by dashed red lines, and its total thickness at different positions is indicated. (C, D) Additional interface between a thick (>1000 layers) bottom flake and a thin (4-7 layers) top flake. Similar potential drops between the domains are observed independent of the thickness of the structures or the substrate: graphite ($SiO_2$) in A (C) respectively.

### f) Dipole moment calculations

To evaluate the dipole moment developing in the system we considered a finite AB stacked hexagonal *h*-BN bilayer model with a surface area of 1.1 nm$^2$ and armchair edges. The flake was initially constructed with uniform B-N bond lengths of 1.446 Å and the edges were saturated by hydrogen atoms with initial B-H and N-H bond lengths of 1.200 Å and 1.020 Å, respectively (Fig. S3A,C). The structures were optimized using the hybrid B3LYP exchange-correlation density functional approximation and the double-ζ polarized 6-31G** Gaussian basis set [40] as implemented in the Gaussian 16 suite of programs [41]. This was followed by refined relaxation adding Grimme's D3 dispersion correction [42] and using the 6-31+G** basis set. Finally, single point calculations were performed on the minimized structures at the B3LYP/6-31+G** and B3LYP/Def2TZVP [43] level of theory. Comparison of the out-of-plane dipole moment components obtained using the three basis sets is provided in Table S1, showing that our results are well converged with respect to basis set size. The value calculated by Def2TZVP was used in Fig. 1C in the main text.

To verify that the flake size used is sufficiently large, we repeated the dipole moment calculation for a bilayer flake with surface area of 2.9 nm$^2$. As seen in Table S1 the obtained values are within 20-25% with those of the smaller flake, indicating that edge effects are relatively small and validating the qualitative value of the results. This is further supported by the Mulliken charge analysis map provided in Fig. S3B, D, showing that most of the charge is located in the bulk region of the flakes.

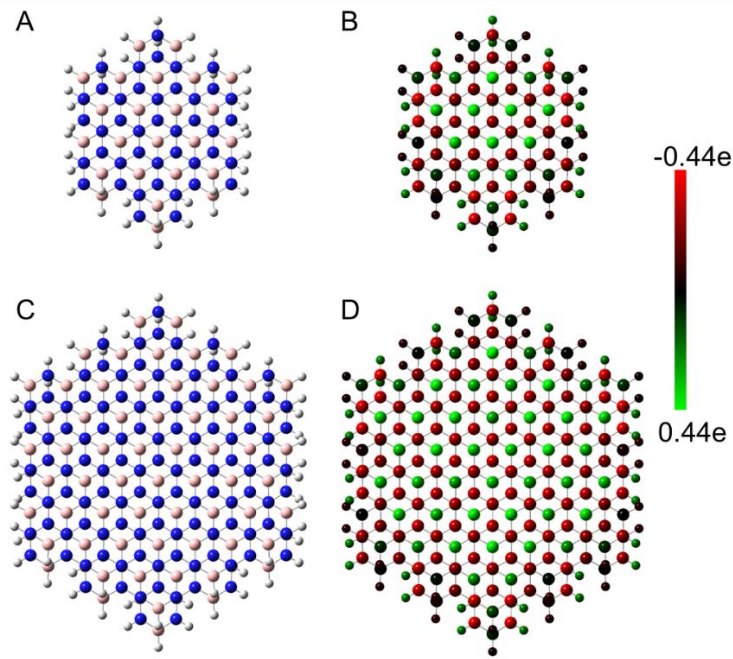

**Fig. S3. Mulliken charge distribution maps;** (A) Top view of the relaxed AB stacked hydrogen terminated finite bilayer $h$-BN flake of 1.1 nm$^2$ contact area. Pink, blue and white spheres represent boron, nitrogen, and hydrogen atoms, respectively. (B) The Mulliken atomic charge map of (A) calculated at the B3LYP/Def2TZVP level of theory. (C) and (D) same as (A) and (B) but for the 2.9 nm$^2$ contact area system, respectively.

**Table S1.** The out-of-plane dipole moments ($p_z$) obtained for two AB stacked bilayer $h$-BN finite flake models.

| Area (nm$^2$) | Dipole moment (Debye/nm$^2$) | | |
|---|---|---|---|
| | 6-31G** | 6-31+G** | Def2TZVP |
| 1.1 | 0.66 | 0.55 | 0.55 |
| 2.9 | 0.52 | 0.45 | 0.45 |

**g) Minimalistic classical cohesive energy model**

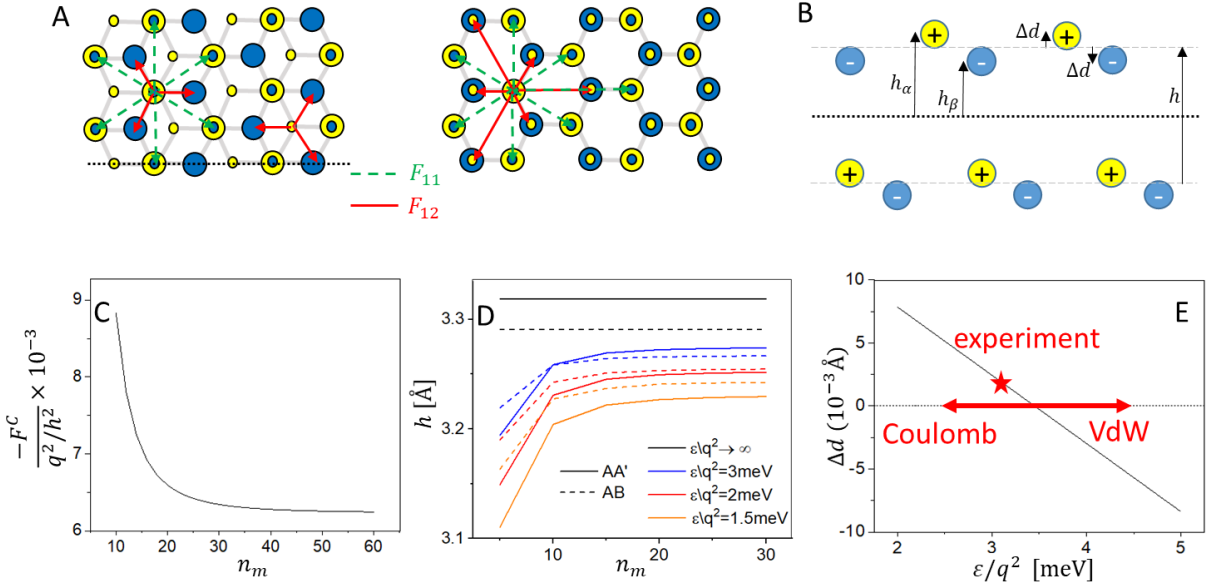

**Fig S4: Classical cohesion model.** (A) Top view of AB (left) and AA' (right) stacked bilayer $h$-BN. The top (bottom) layer atoms are marked by small (large) circles. Lattice sites that participate in the forces $F_{11\,(12)}$ in Eq. (S9) are marked by dashed green (solid red) arrows, respectively. Similar to AA', the eclipsed atoms in AB experience both forces, however, the hollow atoms include only $F_{12}$ (twice). Also note the zero Coulomb force in the latter case due to opposite charges of yellow / blue sites. (B) Cross section of AB stacked bilayer $h$-BN along the dotted black line marked in (A). $\alpha(\beta)$ indicates hollow (eclipsed) sites respectively. (C) Convergence of the total interlayer Coulomb force for $h = 3.3$Å with the number of Bravais lattice vectors ($n_m$) in the summation in Eq. (S9). (D) Inter-layer spacing ($h$) calculated for AA' (solid lines) and AB (dashed lines) stacked bilayer $h$-BN, for different fixed values of cohesion / Coulomb ratio $\varepsilon/q^2$. Black lines with $q = 0$ correspond to graphite and show smaller $h$ for AB than AA' stacking as expected, while orange lines with $\frac{\varepsilon}{q^2} = 1.5$meV show the opposite, as expected for $h$-BN. (E) Intra-layer displacement (marked in B) as a function of $\varepsilon/q^2$. The estimated value from the experiment is marked by a red star, suggesting $\varepsilon/q^2$~3 meV.

Our minimalistic model provides a classical estimate of the out-of-plane polarization of the AB bilayer interface treating the boron and nitrogen atoms as point charges (see Fig. S4A,B), interacting via Pauli and van der Waals (VdW) forces (described by the Lennard-Jones (LJ) potential), and Coulomb interactions. The total interlayer energy is written as follows:
$$E = \frac{1}{2}\sum_{i,j}\left[4\varepsilon\left(\left(\frac{\sigma}{r_{ij}}\right)^{12} - \left(\frac{\sigma}{r_{ij}}\right)^{6}\right) + \frac{q_i q_j}{r_{ij}}\right], \quad (S3)$$

where $\varepsilon$ is the cohesive energy with $\sigma \equiv 3.3$ Å. We note that realistic models of $h$-BN should take atom specific $\varepsilon$ and $\sigma$ values. Here, however we are interested in a qualitative description of the system and hence, for simplicity, we limit the treatment for uniform parameter values. The differences in electronegativity of the boron and nitrogen atoms are effectively taken into account by assigning dimensionless partial charges located at the nuclear centers $q = \pm q_i/e$ for $i \in B, N$ respectively. The parameter $q^2/\varepsilon\sigma$ controls the relative strength between the Coulomb and LJ interactions. As we will demonstrate, this competition determines the sign of the polarization at the AB interface. We denote by α the atomic sites in one layer that reside above hexagon centers in the other layer (termed herein as hollow sites). Correspondingly, $\beta$ denotes atomic sites in one layer that reside above oppositely charged sites on the adjacent layer

(termed herein as eclipsed sites). In each layer we use $h_\alpha$ or $h_\beta$ to denote vertical heights of $\alpha$ and $\beta$ atomic sites, measured with respect to the midplane of the AB interface. To compute the polarization, we minimize the classical energy with respect to $h_\alpha, h_\beta$, via an approximate two-step protocol:

1. First, we set $h_\alpha = h_\beta = h/2$ and minimize the interaction energy with respect to $h$.
2. Then we allow for finite relative vertical motion of B-N pairs around the optimal $h$ value $2\Delta d = h_\alpha - h_\beta$, which generates the polarization. Note that no lateral atomic motion is allowed.

**Optimal interlayer spacing at the AA' stacking mode**

As a reference, we first consider two $h$-BN layers in the AA' stacking configurations with $h_\alpha = h_\beta = h/2$. The total force per atom is:
$$F_{AA'}(h) = -\frac{dE_{AA'}(h)}{dh} = F^{LJ}_{AA'} + F^{C}_{AA'}. \qquad (S4)$$

The Coulomb contribution can be written as $F^{C}_{AA'} = -F^{C}_{11} + F^{C}_{12}$ (with $F^{C}_{11}, F^{C}_{12} > 0$), where
$$F^{C}_{11}(h) = \sum_{\vec{R}_{11}} \frac{e^2 h}{(\vec{R}_{11}^2 + h^2)^{3/2}} \;, \quad F^{C}_{12}(h) = \sum_{\vec{R}_{12}} \frac{e^2 h}{(\vec{R}_{12}^2 + h^2)^{3/2}}. \qquad (S5)$$

Here, $\vec{R}_{11} = \vec{R}_{n1,n2}$ denote in plane lattice vectors connecting equivalent atoms, namely Bravais lattice vectors, and $\vec{R}_{12} = \vec{R}_{n1,n2} - \hat{x}R_0$ denote in plane lattice vectors connecting inequivalent atoms, where $\hat{x}R_0$ is a vector connecting nearest-neighbors. The corresponding Bravais lattice vectors of the honeycomb lattice are given by $\vec{R}_{n_1,n_2} = n_1 \vec{R}_1 + n_2 \vec{R}_2$, and $\vec{R}_{1,2} = R_0(\frac{3}{2}, \pm\frac{\sqrt{3}}{2})$, with $R_0 = 1.4$ Å. Quick convergence of $F^{C}_{AA'}$ is guaranteed if for any pair of integers $n_1, n_2$ the term $F^{C}_{11}(h)$ is combined with $F^{C}_{12}(h)$ calculated for $-n_1, -n_2$ and the sums are taken over the range $-n_m \leq n_1, n_2 \leq n_m$ with sufficiently large $n_m$. The force then converges as $1/n_m^2$ (not shown) in Fig. S4C. Notably, the attractive force $e^2/h^2$ associated with a single vertical bond is strongly suppressed due to the alternating charges within the layer and the small ratio $R_0/h$. This reduces the bare Coulomb interlayer energy $\frac{e^2}{h} \sim 4.3$eV into the meV regime, comparable with the VdW scale $\varepsilon$ [44].

Similarly, the LJ force can be split as $F^{LJ}_{AA'} = F^{LJ}_{11} + F^{LJ}_{12}$, where
$$F^{LJ}_{11}(h) = 4\varepsilon \sum_{\vec{R}_{11}} \left( \frac{12\sigma^{12} h}{(\vec{R}_{11}^2 + h^2)^7} - \frac{6\sigma^6 h}{(\vec{R}_{11}^2 + h^2)^4} \right), \quad F^{LJ}_{12}(h) = 4\varepsilon \sum_{\vec{R}_{12}} \left( \frac{12\sigma^{12} h}{(\vec{R}_{12}^2 + h^2)^7} - \frac{6\sigma^6 h}{(\vec{R}_{12}^2 + h^2)^4} \right). \qquad (S6)$$

The zero-force condition yields the optimal interlayer distance $h$, marked in Fig. S4D by solid lines. As shown $h$ decreases upon decreasing the Lenard-Jones energy scale $\varepsilon$ with respect to the Coulomb energy.

We note that a reasonable approximation for $F^{LJ}_{AA'}(h)$, for small $R_0/h$, consists of treating the particles as a uniform mass distribution, i.e. replacing the sum over $\vec{R}$ by integration, yielding
$$F^{LJ}_{continuum}(h) = \lim_{R_0/h \to 0} F^{LJ} = \varepsilon \frac{32\pi}{3\sqrt{3}R_0^2} h \left( \frac{\sigma^{12}}{h^{12}} - \frac{\sigma^6}{h^6} \right). \qquad (S7)$$

**Optimal interlayer spacing at the AB stacking mode**

Now consider two $h$-BN layers at the AB stacking configuration. The corresponding interlayer force can be written as $F_{AB} = F^{LJ}_{AB} + F^{C}_{AB}$. The unit cell consists of two types of atomic sites, one type where atoms of the two layers reside atop of each other (eclipsed) and the other type where an atom of one layer resides atop a hexagon center of the other layer (hollow sites). Note that the Coulomb contribution of the hollow sites vanishes due to symmetry considerations. Hence, we only have the Coulomb contribution from the eclipsed atomic sites. Since the latter have the exact same configuration in the AA' and AB stacking modes (see Fig. S4A) the overall Coulomb force contribution per atom in the AB stacking mode is half of that in the AA' mode, $F_{AB}^C = \frac{1}{2} F_{AA'}^C$.

For the VdW part, similar to the AA' stacking case, the eclipsed atomic sites give $F_{11}^{LJ} + F_{12}^{LJ}$ of Eq. S10, whereas the hollow atomic sites give $2F_{12}^{LJ}$ due to the unique symmetry of the AB stacked bilayer hexagonal lattice. Therefore, in total we obtain for the LJ force contribution per atom that $F_{AB}^{LJ} = \frac{F_{11}^{LJ}+F_{12}^{LJ}}{2} + \frac{2F_{12}^{LJ}}{2} = \frac{F_{11}^{LJ}}{2} + \frac{3F_{12}^{LJ}}{2}$. The zero-force condition yields the optimal AB stacking interlayer distance, marked by the dashed gray line in Fig. S4D. Similar to the case of AA' stacking mode, $h$ decreases upon decreasing $\varepsilon/q^2$. Note however, that while $h$ (AA') > $h$ (AB) for large $\varepsilon/q^2$ the situation is inverted for $\frac{\varepsilon}{q^2} \lesssim 1.5$ meV. Specifically, when $q \to 0$ our model corresponds to the case of graphite with optimal AB stacking mode.

**Relative vertical displacement**

On top of the interlayer spacing, we now allow a small opposite motion of the hollow site ($\alpha$) and eclipsed site ($\beta$) atoms:
$$h_\alpha = h/2 + \Delta d, \quad h_\beta = h/2 - \Delta d, \quad \text{(S8)}$$
as marked in Fig. S4B. We now analyze the total energy of the system as a function of $\Delta d$, via the Harmonic approximation $E \cong const + \left(\frac{dE}{d(\Delta d)}\right)_{\Delta d=0} \Delta d + \frac{1}{2}\left(\frac{d^2E}{d(\Delta d)^2}\right)_{\Delta d=0} \Delta d^2$ around $\Delta d = 0$. The assumption that $\Delta d \ll R_0$ can be justified by the experimental estimate of $\Delta d \sim 10^{-3}$ Å (see main text). Our simple model considered here, similar to an Einstein model for lattice vibrations, assumes that the $\Delta d$ coordinates are independent Harmonic oscillators with spring constant $K_{\Delta d} = \frac{d^2E}{d(\Delta d)^2}$. Crucially, $\Delta d = 0$ is not a minimum due to the reduced symmetry of the AB interface, with alternating eclipsed and hollow sites, which imposes a finite normal force (per atom) of:
$$F_{\Delta d} = -\frac{dE}{d(\Delta d)} = F_{\Delta d}^C + F_{\Delta d}^{LJ}, \quad \text{(S9)}$$
where $F_{\Delta d}^C$ and $F_{\Delta d}^{LJ}$ are defined as the relative displacement force contributions of the Coulomb and LJ terms. Note that $F_{\Delta d}$ is the difference between the forces acting on the eclipsed versus the hollow sites.

Since, for $\Delta d = 0$, hollow site atoms see a locally charge-neutral configuration on the other layer the Coulomb part of this linear force originates only from the eclipsed atomic sites. For the latter, the atom in the upper layer is attracted to the one residing exactly below it in the other layer. However, it is repelled by its next nearest neighbors in the other layer, and so on. When performing the entire lattice sum for the eclipsed site we obtain that the overall force is always attractive:
$$F_{\Delta d}^C = \frac{F_{11}^C(h) - F_{12}^C(h)}{2} > 0. \quad \text{(S10)}$$

On the other hand, the LJ potential, which can be written as the difference between the hollow site $\left(2F_{12}^{LJ}(h)\right)$ and the eclipsed site $\left(F_{11}^{LJ}(h) + F_{12}^{LJ}(h)\right)$ contributions yields a repulsive force per atom near the equilibrium interlayer distance:

$$F_{\Delta d}^{LJ} = \frac{2F_{12}^{LJ}(h)}{2} - \frac{F_{11}^{LJ}(h) + F_{12}^{LJ}(h)}{2} = \frac{F_{12}^{LJ}(h) - F_{11}^{LJ}(h)}{2} < 0. \tag{S11}$$

This signifies that at the equilibrium interlayer distance, the eclipsed site contribution is more repulsive than that of the hollow site counterpart, mainly due to the fact that the eclipsed atoms are forced to reside within the steep Pauli repulsion wall side of their pairwise interaction.

Overall, from Eqs. S15 and S16 our crude estimate yields

$$\Delta d \sim \frac{F_{\Delta d}}{K_{\Delta d}} = \frac{\left(F_{11}^C(h) - F_{12}^C(h)\right) + \left(F_{12}^{LJ}(h) - F_{11}^{LJ}(h)\right)}{2K_{\Delta d}}. \tag{S12}$$

$K_{\Delta d}$ in Eq. (S17) can be evaluated from the model parameters (see next section). Nevertheless, for simplicity we take it to be equal to the corresponding out-of-plane force constant in graphite $K_{\Delta d} \sim 5$ N/m [45].

The resulting relative displacement is plotted in Fig. S4E versus $\varepsilon/q^2$. Crucially, it changes sign when $\varepsilon_{\text{eff}} = \frac{\varepsilon}{q^2} \sim$ 3.5 meV. The experimentally measured voltages indicate that $\Delta d$ in AB stacked bilayer $h$-BN is of the order of $10^{-3}$ Å (see main text) and our DFT calculations indicate that it is positive suggesting that $\frac{\varepsilon}{q^2} \sim 3$ meV (see Fig. S4E), similar to expected values [44]. Notably, other layered materials, which possess different effective $\varepsilon$ and or $q$ values may show different quantitative and even qualitative polarization.

Finally, it should be noted that our simplistic classical approach is sufficiently flexible to allow to study additional effects such as the dependence on the number of layers as well as external perturbations like pressure or electric field, as well as an additional in-plane component of the polarization, which we leave for future work.

**Analytic estimate of the normal spring constant**

To determine $K_{\Delta d} = \frac{d^2 E}{d(\Delta d)^2}$, we note that it has contributions from the interlayer LJ and Coulomb forces, as well as from the intra-layer forces. Its interlayer LJ contribution is obtained from the corresponding contribution to the energy $E^{LJ}(\Delta d) = \frac{1}{2}\left[V_{11}^{LJ}(h) + V_{12}^{LJ}(h - 2\Delta d)\right] + \frac{1}{2}\left[V_{12}^{LJ}(h) + V_{12}^{LJ}(h + 2\Delta d)\right]$. The first (second) term represents the interaction of atoms at eclipsed (hollow) sites, and $V_{11(12)}^{LJ}(h) = 4\varepsilon \sum_{\vec{R}_{11(12)}} \left(\frac{\sigma^{12}}{\left(\vec{R}_{11(12)}^2 + h^2\right)^6} - \frac{\sigma^6}{\left(\vec{R}_{11(12)}^2 + h^2\right)^3}\right)$. One can obtain a good approximation for the spring constant $\left(\frac{d^2 E}{d(\Delta d)^2}\right)_{\Delta d=0}$ by replacing the sums $\sum_{\vec{R}_{11(12)}}$ by integrals, as in $F_{continuum}^{LJ}(h)$. This procedure yields

$$K_{\Delta d}^{LJ} = \frac{64\pi}{\sqrt{3}R_0^2}\varepsilon. \tag{S13}$$

For $\varepsilon$=3meV this yields a spring constant of $K_{\Delta d}^{LJ} = 3$ N/m. While additional contributions are expected from intra-layer interactions, as well as Coulombic inter-layer interactions, this value is comparable to the measured out-of-plane force constant in graphite $K_{\Delta d} \sim 5$ N/m [45]. Since Eq. S13 captures the order of magnitude of $K_{\Delta d}$, we can use $K_{\Delta d} =$

$\frac{64\pi}{\sqrt{3}R_0^2}\varepsilon \cdot a$, with a factor of order unity $a \sim \frac{5}{3}$, and obtain an approximate expression for the relative displacement, see

Eq. (S17), fully in terms of our model's parameters, $\Delta d \sim \frac{F_{\Delta d}}{K_{\Delta d}} = \frac{\sqrt{3}R_0^2}{32\pi} \frac{\left(F_{11}^C(h) - F_{12}^C(h)\right) + \left(F_{12}^{LJ}(h) - F_{11}^{LJ}(h)\right)}{4\varepsilon}$.

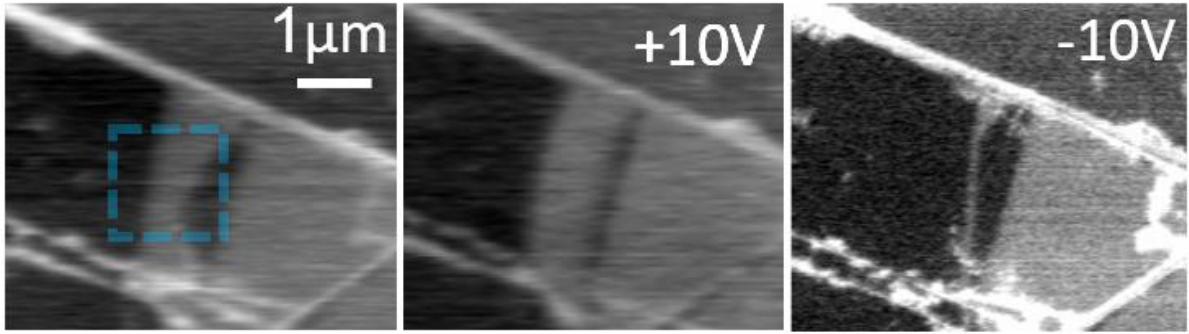

**Fig. S5. Additional example of domain-wall sliding due to biased tip scans.** Consequent KPFM images of the same flake location (from left to right-hand side). Between the images a biased tip (±10 V) was scanned above the region marked by a blue square. Positive tip bias resulted in domain-wall motion that increased the white domains area over the black domains and vice versa.